\documentclass[a4paper]{article}
\usepackage{amsmath}
\usepackage{latexsym}

\begin{document}
\title{New Formulation of Massive Spin Two Field Theory and Solution of vDVZ Discontinuity Problem}
\author{Ki-Ichiro~SATO\thanks{e-mail address: kisato@rs.kagu.tus.ac.jp}
}
\date{Faculty of Industrial Science and Technology,\\
Tokyo University of Science, \\
Oshamanbe-cho, Hokkaido, 049-3514, Japan}
\maketitle
\abstract{We present a new massive spin two field theory which has smooth massless limit based on BRS formalism.
Our model contains a parameter $a$ which extendes Fierz-Pauli's mass term.
Although redundant scalar ghost exists in our model,
 physical degree of freedom is still five for the sake of existence of 
the  Deser-Waldron type gauge symmetry.
Origin of vDVZ discontiunity can be studied at the Fierz-Pauli limit $a \rightarrow 1$.
}

\section{Introduction}
\par
It is difficult to formulate massive field theories with higher spin, 
which have a continuos massless limit,
because, unitary irreducible representations of Poincar\'e group are independently given for massless and massive cases.
Once one requires manifest covariance in field theory,
one have to introduce redundant degree of freedom,
and after quantization procedure one has to remove this
redundant degree of freedom.
This condition is given by Bergman-Wigner equation.
For spin two case, this equation is known as Fierz-Pauli equation and Fierz and Pauli presented a Lagnangian field theory (Fierz-Pauli theory) \cite{FierzPauli}.

However, as is well known, second quantized version of Fierz-Pauli theory
has a discontinuity in massless limit (vDVZ discontinuity) \cite{vDVZ}.
vDVZ discontiunity consist in two part;
One is inperfect spin two projection operator, which brings us incompatible
infrared property for the motion of the planet \cite{BoulwareDeser72},
and anothor is $1/m^2$ singurarity which
brings undesibable high energy behavior of the propagator which is similar to Proca field.
The former is serius problem.

The ghost and tachyon free Langangian theory is nothing but the Fierz-Pauli theory\cite{Nieuwenhuizen73}.
So, it seems that vDVZ discontiunity is unavoidable.
However, when one adandons ghost free condition, situation is changed.
The problem of vDVZ discontiuity had been solved partially by Kimura~\cite{Kimura76}. 
He presented general theories, which is one parameter extension of Fierz-Pauli's mass term, in operator formalism.
His theory has smooth massless limit to linearized gravity and no $1/m^2$ singurarity. 
Unfortunately, Kimura did not present subsidiary condition for the physical subspace in order to assure unitarity, and did not show that physical degree of freedom is five.

In this paper, we would like to present a
new formulation of massive spin two field theory, which is different from Kimura.
We start with the same extended mass term.
In order to ensure off-shell constraint equation for six degree of freedom,
we utilize BRS procedure.
After BRS procedure is done, we find local gauge symmetry which removes one physical degree of freedom.

\section{general tachyon free massive symmetric tensor field with a scalar ghost}
We start with a Lagrangian $\mathcal{L}_{EH}^{(2)}$ as a kinetic term for the symmetric tensor field $h_{\mu\nu}$;
\begin{equation}
\mathcal{L}_{EH}^{(2)} = \frac{1}{2}[\partial_\lambda h^{\mu\nu}\cdot\partial^\lambda h_{\mu\nu}-\partial_\lambda h\cdot\partial^\lambda h -2\partial_\mu h^{\mu\lambda}\cdot\partial^\nu h_{\nu\lambda} + 2 \partial^\mu h_{\mu\lambda}\cdot\partial^\lambda h].
\label{eq:linearlizedLagrangian}
\end{equation}
This Lagrangian is obtained from the Einstein-Hilbert Lagrangian $\mathcal{L}_{\mathrm{EH}} = \frac{1}{\kappa}\sqrt{-g}R $ by taking weak field appoximation around flat Minkowski background $\eta_{\mu\nu}$;
\[
g_{\mu\nu} = \eta_{\mu\nu} + \sqrt{2\kappa} h_{\mu\nu}.
\]
Here, $\kappa$ is Einstein's gravitational constant.
We represent the space-time dimension as $d$.
We currently have interest in the 4 dimensional case,
 but we need generic $d$ dimension for the detailed discussion.

Next, we adopt a mass term
\begin{equation}
{\cal L}_\mathrm{m} = -\frac{m^2}{2}[h^{\mu\nu}h_{\mu\nu}-ah^2],
\end{equation}
which is different from Fierz-Pauli's mass term.
Here, $h$ is a scalar field which is defined by trace part of $h_{\mu\nu}$, $h = \eta^{\mu\nu}h_{\mu\nu}$.
This is one-parameter extension of Fierz-Pauli mass term.
In $a=1$ case, the mass term reduces to Fierz-Pauli mass term.

The equation of motion of the system is obtained as follows;
\begin{eqnarray}
(\Box+m^2)h_{\mu\nu} - \eta_{\mu\nu}(\Box+am^2)h &-& \partial_\mu\partial^\lambda h_{\lambda\nu} - \partial_\nu\partial^\lambda h_{\mu\lambda} \nonumber\\
&+& \partial_\mu\partial_\nu h + \eta_{\mu\nu}\partial^\lambda\partial^\rho h_{\lambda\rho} = 0.
\label{eq:equationOfMotionOfTypeI}
\end{eqnarray}
Divergence of (\ref{eq:equationOfMotionOfTypeI}) becomes
\begin{equation}
m^2(\partial^\mu h_{\mu\nu} - a\partial_\nu h) = 0. 
\end{equation}
If $m \not= 0$, ten componets of $h_{\mu\nu}$ are not independent each other because
there are four constraints;
\begin{equation}
\partial^\lambda h_{\mu\nu} = a\partial_\nu h.
\label{eq:constraintOfTypeI}
\end{equation}
Hence, there remains six degree of freedom in this theory.

Utilizing (\ref{eq:constraintOfTypeI}), we derive an equation of motion for $h_{\mu\nu}$ as follows;
\begin{equation}
(\Box+m^2)h_{\mu\nu} - \eta_{\mu\nu}\Big[ (1-a)\Box + am^2\Big] h + (1-2a)\partial_\mu\partial_\nu h = 0. \label{eq:equationOfMotion2}
\end{equation}
Trace part of this equation becomes
\begin{equation}
(2-d)(1-a) \Box h + (1-da)m^2 h = 0.
\label{eq:TracePartOfEquationOfMotion}
\end{equation}
If $d=2$ and/or $a=1$ are satisfied, the eq. (\ref{eq:TracePartOfEquationOfMotion}) 
 implies a traceless condition, $h = 0$.
This is familiar in Fierz-Pauli theory.

In the case when both $d\not= 2$ and $a\not= 1$ are satisfied, 
we rewrite the eq. (\ref{eq:TracePartOfEquationOfMotion}) as follows;
\begin{equation}
(\Box + M^2 ) h = 0.
\end{equation}
Here, the mass $M$ of trace mode $h$ is defined by
\begin{equation}
M^2(a,d) = \frac{1-da}{2(a-1)}m^2,
\end{equation}
which depends on a parameter $a$ and the space-time dimension $d$.
In general, $M^2$ can take negative value,
and $h$ may be tachyon.
Tachyon free condition is given by
\begin{equation}
\frac{1}{d} \le a < 1.     \label{eq:TachyonFreeCondition}
\end{equation}
In 4-dimensional space-time, tachyon free condition is
$1/4 \le a < 1$.

For the later convenience, we define 
the traceless symmetric tensor field $\bar{h}_{\mu\nu}$
as follows;
\[
\bar{h}_{\mu\nu} = h_{\mu\nu} - \frac{1}{d}\eta_{\mu\nu}h,
\]
Using this notation, we obtain the equation of motion as follows;
\begin{eqnarray}
(\Box+M^2)h &=& 0, \\
(\Box+M^2)(\Box+m^2)\bar{h}_{\mu\nu} &=& 0, \\
\partial^\mu \bar{h}_{\mu\nu} &=& (a-\frac{1}{d})\partial_\nu h.
\end{eqnarray}
The trace mode $h$ has a oppesite signed kinetic term.
So, our theory contains a scalar ghost and multi-mass particles.

It is to be pointed out that there are three special cases in our theory.
\begin{description}
\item[(i) Fierz-Pauli limit ($a \rightarrow 1$)] \mbox{} \\
In general, we consider $a\not= 1$.
Here, we consider the $a \rightarrow 1$ limit.
It is pointed out that $h$ have an infinite mass.
We call this case Fierz-Pauli limit.
At this limit, $M^2$ goes to infinity, hence the trace mode $h$ does not propagate.
But, the field $h$ does not identically vanish, 
and total degree of freedom is still six, not five.
This is crutial point to understand vDVZ discontinuity.

\item[(ii) $a=1/2$ case] \mbox{} \\
In general, the theory contains multi-mass modes.
In $a=1/2$ case, the theory reduces to a simple mass field theory.
In this case, $M^2$ becomes $m^2$, and the scalar mode $h$ satisfies a
single mass equation of motion;
\[
(\Box+m^2)h = 0.
\]
One may expect that $h_{\mu\nu}$ is massive dipole field.
However, $h_{\mu\nu}$ also satisfies,
\begin{equation}
(\Box+m^2)h_{\mu\nu} = 0. \label{eq:}
\end{equation}
Moreover, constraint (\ref{eq:constraintOfTypeI}) turns out to be
\begin{equation}
\partial^\mu h_{\mu\nu} = \frac{1}{2} \partial_\nu h. \label{eq:constraintAisUnity}
\end{equation}
The eq. (\ref{eq:constraintAisUnity}) coinsides with de Donder condition;
\begin{equation}
\partial_\mu (\sqrt{-g} g^{\mu\nu} ) = 0.
\end{equation}
In massless theory, de Donder condition is often taken as a gauge condition.
In de Donder condition and Feynmann-like gauge, massless graviton propagator reduces
into simple form;
\[
\mathcal{F.T.} <T^{*}h_{\mu\nu}h_{\alpha\beta}>
= \frac{1}{p^2}\Big[ \frac{1}{2}\eta_{\mu\alpha}\eta_{\nu\beta}
+\frac{1}{2}\eta_{\mu\beta}\eta_{\nu\alpha}
- \frac{1}{2} \eta_{\mu\nu}\eta_{\alpha\beta} \Big].
\]
If one consider whether massless limit will coinside with the massless theory
with de Donder condition, or not, one has to take $a=1/2$, not $a=1$.
In other words, massless limits of the Fierz-Pauli theory does not coinside
with massless theory with de Donder condition.
This is an important result.

\item[(iii) $a=1/d$ case] \mbox{} \\
Lastly, we study $a=1/d$ case.
In this case, $h$ becomes massless, because the equation of motion is
given by
\[
\Box h = 0.
\]
From this result, $h_{\mu\nu}$ satisfies a equation of motion;
\[
\Box(\Box+m^2)h_{\mu\nu} = \Box(\Box+m^2)\bar{h}_{\mu\nu} = 0.
\]
The traceless part satisfies massless equation of motion turn out to be 
traceless-transverse condition;
\[
\partial^\mu \bar{h}_{\mu\nu}=0.
\]
In $a=1/d$, the Lagrangian is invariant under the scale transformation $h_{\mu\nu} \rightarrow \lambda h_{\mu\nu}$.
Massless modes concerns with this symmmetry.
In Ads and dS space-time, 
strange partially massless state exists in spin two field theory \cite{DeserWaldron01,DolanNappiWitten}.
Massless mode in our model may be related to AdS and/or dS case.
\end{description}

In order to study the vDVZ discontinuity of the theory,
we shall calculate two point function.
This is given by
\begin{eqnarray*}
&&\mathcal{F.T.} <\mathrm{T}^{*}h_{\mu\nu}h_{\alpha\beta}>\\
&=& \frac{1}{p^2-m^2}\Big[ \frac{1}{2}\eta_{\mu\alpha}\eta_{\nu\beta}
+\frac{1}{2}\eta_{\mu\beta}\eta_{\nu\alpha}
- \frac{1}{d-2} \eta_{\mu\nu}\eta_{\alpha\beta} \\
&&- \frac{1}{2m^2}(\eta_{\mu\alpha}p_\nu p_\beta + \eta_{\mu\beta}p_\nu p_\alpha + \eta_{\nu\alpha}p_\mu p_\beta + \eta_{\nu\beta}p_\mu p_\alpha ) \Big] \\
&& + \frac{(2a-1)}{(p^2-m^2)(p^2-M^2)} \Big[ \frac{1}{(d-2)^2(a-1)}m^2 \eta_{\mu\nu}\eta_{\alpha\beta}\\
&& + \frac{1}{(d-2)(a-1)}(\eta_{\mu\nu}p_\alpha p_\beta + \eta_{\alpha\beta}p_\mu p_\nu )
+ \frac{1}{(a-1)m^2} p_\mu p_\nu p_\alpha p_\beta \Big].
\end{eqnarray*}
Spin 2 structure is found in first three terms.
If we set $d$ to be space-time dimension four, correct factor $(1/2)$ realizes.
Then, there is no vDVZ discontinuity of spin 2 structure for any value of $a$ 
in this model.

\section{Physical degree of freedom and gauge symmetry}
Next, we shall consider physical degrees of freedom in this theory.
One cannot solve constraint for $h_{\mu\nu}$ represented by (\ref{eq:constraintOfTypeI}) directly.
In order to assure (\ref{eq:constraintOfTypeI}) in off shell,
we shall adopt gauge condition as follows;
\begin{equation}
\partial^\mu h_{\mu\nu} - a\partial_\nu h = 0, \label{eq:gaugeCondition}
\end{equation}
and, we introduce an BRS transformation which correspondes to the vector gauge transformation;
\begin{equation}
\delta h_{\mu\nu} = \partial_\mu C_\nu + \partial_\nu C_\mu, \ \ 
\delta h = 2 \partial^\mu C_\mu, \ \ 
\delta C_\mu =0, \ \ 
\delta \bar{C}^\mu = iB^{\mu}, \ \ 
\delta B^{\mu} =0.
\end{equation}
Here, $C_\mu$ and $\bar{C}^\nu$ are FP ghosts, 
$B^\mu$ is the Nakanishi-Lautrup field.

We give an BRS invariant Langargian under the gauge condition (\ref{eq:gaugeCondition}) as follow;
\begin{eqnarray}
\mathcal{L}_{\mathrm{BRS}}^{\mathrm{V}} &=& -i \delta [ \bar{C}^{\nu}(-1)(\partial^\mu h_{\mu\nu}-a\partial_\nu h + \frac{1-a}{2m^2}(\alpha_1\Box B_\nu + \alpha_2 \partial_\nu \partial_\mu B^\mu) ] \nonumber \\
&=& (-1)(\partial^\mu h_{\mu\nu}-a\partial_\nu h)B^\nu - \frac{1-a}{2m^2}\left(\alpha_1 (\partial_\mu B^\nu)^2 + \alpha_2 (\partial_\mu B^\mu)^2 \right) \nonumber \\
&& +i \bar{C}^\nu (\Box C_\nu + (1-2a)\partial_\nu\partial^\mu C_\mu).
\end{eqnarray}
We split this Langangian into two parts.
One is a gauge fixing term $\mathcal{L}_{\mathrm{GF}}^{\mathrm{V}}$ which does not contain FP ghosts,
and anothor is a FP ghosts term $\mathcal{L}_{\mathrm{FP}}^{\mathrm{V}}$;
\begin{equation}
\mathcal{L}_{\mathrm{GF}}^{\mathrm{V}} = (-1)(\partial^\mu h_{\mu\nu}-a\partial_\nu h)B^\nu - \frac{1-a}{2m^2}\left(\alpha_1 (\partial_\mu B^\nu)^2 + \alpha_2 (\partial_\mu B^\mu)^2 \right),
\end{equation}
\begin{equation}
\mathcal{L}_{\mathrm{FP}}^{\mathrm{V}} = i \bar{C}^\nu (\Box C_\nu + (1-2a)\partial_\nu\partial^\mu C_\mu).
\end{equation}
As $B^\mu$ is dynamical in the case of $a\not= 1$,
physical degree of freedom is still six via native counting $ 10(h_{\mu\nu})+ 4(B^\mu) - 4\times 2(C_\mu,\bar{C}^\nu) = 6$ , not five.
So, we have to find the gauge invariance which remove one physical degree of the freedom.
The common factor $1-a$ of the two gauge paramers, $\alpha_1$ and $\alpha_2$,
is chosen for the later convenience.

In the following, we will consider the mass term and gauge fixing term;
\[
\mathcal{L}_m + \mathcal{L}_{\mathrm{GF}}^{V}.
\]
Here, we introduce the gauge transformation which
is some kind of Deser-Waldron's gauge transformation \cite{DeserWaldron01};
\[
\delta h_{\mu\nu} = \frac{1}{m^2}(\partial_\mu\partial_\nu - b\eta_{\mu\nu}\Box ) \sigma(x), \quad \delta h = \frac{1-bd}{m^2}\Box \sigma.
\]
Here, $\sigma(x)$ is an arbitrary function of the space-time,
and $b$ is a parameter.

It is interesting in the case when the $b$ is taken to be zero;
\begin{eqnarray}
\delta h_{\mu\nu} &=& \frac{1}{m^2}\partial_\mu\partial_\nu \sigma(x),
\quad \delta h = \frac{1}{m^2}\Box \sigma, \nonumber \\
\delta B^\mu &=& \partial^\mu \sigma(x), \quad 
\delta C_\mu = \delta \bar{C}^\nu = 0. \label{eq:DeserWaldronTransformation}
\end{eqnarray}
Under the gauge transformation (\ref{eq:DeserWaldronTransformation}) for $a\not=1 $ case,
If 
\begin{equation}
\alpha_1+\alpha_2 = 0,
\end{equation}
is satisfied, $\mathcal{L}_m + \mathcal{L}_{\mathrm{GF}}$ is invariant.
In the case of $a=1$, $\mathcal{L}_m + \mathcal{L}_{\mathrm{GF}}$ is invariant under the condition $\alpha_1=\alpha_2=0$.
Existence of the factor $1-a$ describes this situation.
Massive neutral vector field theory which has smooth massless limit presented by Nakanishi \cite{Nakanishi72}.
$a=1$ case is corresponding to Nakanishi's procedure because our $B^\mu$ is not dynamical.
For $a\not= 1$, our formulation is some kind of extension to Nakanishi's formulation.

Kinetic term $\mathcal{L}_{\mathrm{EH}}^{(2)}$ and FP ghosts term are also invariant, hence total Lagrangian is invariant under the gauge transformation (\ref{eq:DeserWaldronTransformation}).

We have found a new gauge symmetry in the system.
Next, we introduce BRS transformation corresponding to 
the gauge transformation (\ref{eq:DeserWaldronTransformation});
\begin{eqnarray}
\delta^\mathrm{S} h_{\mu\nu} &=& \partial_\mu\partial_\nu C,
\quad \delta^\mathrm{S} h = \Box C, \quad \delta^\mathrm{S} B^\mu = \partial^\mu C, \nonumber \\
\delta^\mathrm{S} C_\mu &=& \delta^\mathrm{S} \bar{C}^\nu = 0, \\
\delta^\mathrm{S} C &=& 0, \quad \delta^\mathrm{S} \bar{C} = iB, \quad \delta^\mathrm{S} B = 0, \nonumber
\end{eqnarray}
$B$ is a Nakanishi-Lautrap field, $C$ and $\bar{C}$ are FP ghosts.
For new fields, BRS transformaion of vector type is defined by
\begin{equation}
\delta B = \delta C = \delta \bar{C} = 0,
\end{equation}

In order to make degree of freedom clearer, we adopt a traceless gauge;
\[
h = 0,
\]
The gauge fixing term and FP ghost term are given by
\begin{equation}
\mathcal{L}_{BRS}^\mathrm{S} = -i\delta^\mathrm{S} \left[ \bar{C} h \right] 
= hB + i \bar{C} \Box C.
\end{equation}
As $B$ is auxiliary field, we can eliminate $h$ in our Lagrangian,
and $C$ and $\bar{C}$ are decouple.
In this gauge condition, physical degree of freedom is five via counting $9(\bar{h}_{\mu\nu})+4(B^{\mu})-4\times 2(C_\mu,\bar{C}^{\nu})= 5$.

For example, we consider the case of $\alpha_1=1$, $\alpha_2=0$ and $a=1/2$.
In this case, equations of motion are given by
\begin{eqnarray}
(\Box+m^2) \bar{h}_{\mu\nu} - \partial_\mu \partial^\lambda \bar{h}_{\lambda\nu}
 - \partial_\nu \partial^\lambda \bar{h}_{\lambda\mu}
 + \eta_{\mu\nu}\partial^\lambda \partial_\rho \bar{h}_{\lambda\rho} && \nonumber\\
-\frac{1}{2}(\partial_\mu B_\nu + \partial_\nu B_\mu-\eta_{\mu\nu}\partial_\lambda B^{\lambda}) - \eta_{\mu\nu} B &=& 0, \\
\partial_\mu \bar{h}^{\mu\nu} -\frac{1}{2m^2}\Box B^\nu &=& 0, \label{eq:subsidiaryCondition} \\
\Box B = \Box C = \Box \bar{C} = \Box C_\mu = \Box \bar{C}^\nu &=& 0. 
\end{eqnarray}
In this derivation, we use $h=0$ and we ignore a trace part of equation of motion for $h_{\mu\nu}$.

In the folloing, we shall consider the case when $C$ and $\bar{C}$ are absent in physical subspace.
Unfortunately, Vector BRS transformation is not usefull, 
becuase there is no vector BRS invariance in the full Lagrangian.
Actually, BRS current $J^\mu$ which is given by
\begin{equation}
J^\mu = B^\lambda\partial^\mu C_\lambda - \partial^\mu  B^\lambda\cdot C_\lambda,
\end{equation}
does not satisfy continuity equation.
Its breaking is expricitly calculated by
\begin{equation}
\partial_\mu J^\mu = 2m^2 \partial_\mu \bar{h}^{\mu\nu}\cdot C_\nu,
\end{equation}
If 
\begin{equation}
\partial_\mu \bar{h}^{\mu\nu} = 0,
\end{equation}
is satisfied, BRS invariance will be recovered.
This is nothing but traceless-transverse condition for unitarity.
Using equation of motion (\ref{eq:subsidiaryCondition}), in order to realize this condition,
$B^\mu$ has to anihilate any state vector in our physical subspace.
Unfortunately, multi-mass nature of $\bar{h}_{\mu\nu}$ complicates this
 analysis, details of additional subsidiary condition is not given here.
This will be done in the separated paper.

\section{Fierz-Pauli limit in general covariant gauge}
We shall briefly study Fierz-Pauli limit in our model.
We here take the general covariant gauge,
\begin{equation}
\partial_\mu B^\mu = 0,
\end{equation}
instaed of $h=0$.
Under this condition, gauge fixing and FP ghost terms are given by
\begin{equation}
\mathcal{L}_{BRS}^{S} = -i\delta [\bar{C}(\partial_\mu B^\mu + \frac{\alpha}{2}B) = \partial_\mu B^\mu\cdot B + \frac{\alpha}{2}B^2 + i \bar{C}\Box C,
\end{equation}
where $\alpha$ is a gauge parameter.
The equations of motion for $h_{\mu\nu}$, $B^\mu$ and $B$ are obtained as follows;
\begin{eqnarray}
\Box h_{\mu\nu} - \eta_{\mu\nu}\Box h - \partial_\mu \partial^\lambda h_{\lambda\nu} - \partial_\nu \partial^\lambda h_{\mu \lambda} && \nonumber \\
+ \partial_\mu \partial_\nu h + \eta_{\mu\nu}\partial^\lambda\partial^\rho h_{\lambda\rho}
+ m^2 h_{\mu\nu} -am^2\eta_{\mu\nu} h  && \nonumber\\
-\frac{1}{2}(\partial_\mu B_\nu + \partial_\nu B_\mu)
+ a \eta_{\mu\nu} \partial_\lambda B^\lambda &=& 0, \label{eq:EqOfMotionh}
\end{eqnarray}
\begin{equation}
\partial^\mu h_{\mu\nu} -a \partial_\nu h - \frac{1-a}{m^2}[\alpha_1\Box B_\nu + (1-\alpha_1)\partial_\nu \partial_\lambda B^\lambda] + \partial_\nu B = 0, \label{eq:EqOfMotionBmu}
\end{equation}
\begin{equation}
\partial_\mu B^\mu + \alpha B = 0, \label{eq:EqOfMotion}
\end{equation}
\begin{eqnarray}
\Box C_\mu + (1-2a) \partial_\mu \partial^\lambda C_\lambda &=& 0, \\
\Box \bar{C}^\mu + (1-2a) \partial^\mu \partial_\lambda \bar{C}^\lambda &=& 0, \\
\Box C = \Box \bar{C} &=& 0,
\end{eqnarray}
By conbinning with the equations of motion, we obtain the equations of motion for arbitray parameters $a$ and $\alpha_1$;
\begin{eqnarray}
\Box^2 (\Box + m^2)(\Box + M^2) h_{\mu\nu} &=& 0, \nonumber\\
\Box (\Box + M^2) h &=& 0, \\
\Box B = \Box C = \Box \bar{C} &=& 0, \nonumber\\
\Box^2 B^\mu = \Box^2 C_\nu = \Box^2 \bar{C}^\lambda &=& 0. \nonumber
\end{eqnarray}

It is to be pointed out that $a=1$ can be taken in our model.
In $a=1$ case, the equations of motion become
\begin{eqnarray}
\Box^2 (\Box + m^2) h_{\mu\nu} &=& 0, \nonumber\\
\Box h = \Box B = \Box C = \Box \bar{C} &=& 0, \\
\Box^2 B^\mu = \Box^2 C_\nu = \Box^2 \bar{C}^\lambda &=& 0, \nonumber
\end{eqnarray}
for the sake of first BRS procedure.
At first, $h$ has infinite mass in the limit $a \rightarrow 1$.
Now, $h$ becomes massless, and it does propagate.
After final BRS procedure, we obtain the theory for $1/d \le a \le 1$.
It is possible to take Fierz-Pauli limit in our model.

\section{Concluding remarks}
We have presented a new massive spin two field theory based on BRS procedure.
Our model contains Fierz-Pauli theory in the special case of $a=1$.
Origin of vDVZ discontinuity strongly depends on the fact that scalar mode $h$ has an infinite mass.
Five physical degree of freedom is assured for the sake of hidden gauge symmetry like Deser-Waldron's one.
While, our model has no tachyon for $1/d \le a \le 1$.
Particle contents is drastically changed when one varies the value of $a$.
Therefore, our theories consists in five different models; (1) $a=1$ model (Fierz-Pauli theory), (2) $a=1/2$ model, (3) $a=1/d$ model, (4) $a \in (1/d,1/2)$ model and (5) $a \in (1/2,1)$ model.
In addition to it, the existence of gauge parameter $\alpha_1$ make our model more complicated.
So, we has presented an explanation for unitaity and correct five physical degree of freedom at special case of $a=1/2$, $\alpha_1=1$, and $h=0$ gauge.
Strict proof of unitarity has to be done earlier.

There are other researches about vDVZ discontinuity than Kimura's work.
$a=1/2$ model has studied by Fronsdal and Heidenreich \cite{FronsdalHeidenreich92}.
They formulated $a=1/2$ model based on exact two BRS symmetry; vector type BRS transformation and scalar type BRS transformation like Weyl transformation.
Their theory has smooth massless limit and no $1/m^2$ singurality.
Inspite of exact two BRS invariance, they counld not show unitarity.
We have an interest in difference between their theory and ours.


\begin{thebibliography}{9}
\bibitem{FierzPauli}
M. Fierz and W. Pauli, Proc. Roy. Soc. London {\bf A173} (1939) 211.
\bibitem{vDVZ}
H. van Dam and M. Veltman, Nucl. Phys. {\bf B22} (1970) 397.\\
V. I. Zakharov, JTEP Lett. {\bf 12} (1970) 312.
\bibitem{BoulwareDeser72}
D. G. Boulware and S. Deser, Phys. Rev. {\bf D6} (1972) 3368.
\bibitem{Nieuwenhuizen73}
P. van Nieuwenhuizen, Nucl. Phys. {\bf B60} (1973) 478.
\bibitem{Kimura76}
T. Kimura, Prog. Theor. Phys. {\bf 55} (1976) 1259; Prog. Theor. Phys. {\bf 55} (1976) 1328.
\bibitem{DeserWaldron01}
S. Deser and A. Waldron, Phys. Lett. {\bf B501} (2001) 134; Phys. Rev. Lett. {\bf 87} (2001) 031601; Nucl. Phys. {\bf B607} (2001) 577; Phys. Lett. {\bf B508} (2001) 347; Phys. Lett. {\bf B513} (2001) 137;\\
S. Deser and B. Tekin, Class. Quan. Grav. {\bf 18} (2001) L171.
\bibitem{DolanNappiWitten}
L. Dolan, C. R. Nappi, and E. Witten, JHEP {\bf 10} (2001) 016.
\bibitem{Nakanishi72}
N. Nakanishi, Phys. Rev. {\bf D5} (1972) 1324; Prog. Theor. Phys. Suppl. , {\bf 51} (1972) 1; ``Quantum Field Theory'' (Baifukan 1975), in Japanese.
\bibitem{FronsdalHeidenreich92}
C. Fronsdal and W. F. Heidenreich, Ann. Phys. {\bf 215} (1992) 51.
\end{thebibliography}
\end{document}